\documentclass[oldversion]{aa}
\usepackage[varg]{txfonts}
\usepackage{color,graphics}
\begin{document}

\title{Can single O stars produce non-thermal radio emission?
}
\titlerunning{Non-thermal emission from single O stars}
\author{S. Van Loo \inst{1,2}, M.C. Runacres \inst{1,3} \and R. Blomme \inst{1}}
\authorrunning{Van Loo et al}
\institute{Royal Observatory of Belgium, Ringlaan 3, B-1180 Brussel, Belgium
\and School of Physics and Astronomy, University of Leeds, Leeds LS2 9JT, UK
\and Vrije Universiteit Brussel, Pleinlaan 2, B-1050 Brussel, Belgium}
\offprints{S. Van Loo,\\ \email{svenvl@ast.leeds.ac.uk}}
\date{Received / Accepted}
\abstract{
We present a model for the non-thermal radio emission from presumably single O 
stars, in terms of synchrotron emission from relativistic electrons accelerated 
in wind-embedded shocks. These shocks are associated with an unstable, chaotic wind. 
The main improvement with respect to earlier models is the inclusion of the radial 
dependence of the shock velocity jump and compression ratio, based on one-dimensional 
time-dependent hydrodynamical simulations. The decrease of the velocity jump and the 
compression ratio as a function of radius produces a rapidly decreasing synchrotron
emissivity. This effectively prohibits the models from reproducing the spectral shape 
of the observed non-thermal radio emission. We investigate a number of ``escape routes" 
by which the hydrodynamical predictions might be reconciled with the radio observations. 
We find that the observed spectral shape can be reproduced by a slower decline of the 
compression ratio and the velocity jump, by the re-acceleration of electrons in many 
shocks or by adopting a lower mass-loss rate. However, none of these escape routes are 
physically plausible. In particular, re-acceleration by feeding an electron distribution 
through a number of shocks, is in contradiction with current hydrodynamical simulations. 
These hydrodynamical simulations have their limitations, most notably the use of 
one-dimensionality. At present, it is not feasible  to perform two-dimensional 
simulations of the wind out to the distances required for synchrotron-emission models. 
Based on the current hydrodynamic models, we suspect that the observed non-thermal radio 
emission from O stars cannot be explained by wind-embedded shocks associated with the 
instability of the line-driving mechanism. The most likely alternative mechanism is 
synchrotron emission from colliding winds. That would imply that all O stars with 
non-thermal radio emission should be members of binary or multiple systems.
}
\maketitle

\section{Introduction}
Hot stars are sources of both thermal and non-thermal radio emission. 
The {\em thermal} radiation is free-free emission from thermal electrons 
in the ionised stellar wind. As all hot stars have winds, they are all
expected to have thermal emission, although it is intrinsically faint 
and can only be detected for the brightest and closest stars.
A significant fraction of hot stars also emits {\em non-thermal} 
radiation, which is generally attributed to synchrotron emission from 
relativistic electrons. 

Wright \& Barlow (\cite{WB75}) developed an elegant formalism for the 
thermal radio emission from hot stars. Assuming a uniform, isothermal, 
spherically symmetric outflow, they showed that the expected flux is of 
the form $F_{\nu} \propto \nu^{\alpha}$, where the spectral index $\alpha=+0.6$.
The same formalism allows a star's mass-loss rate to be derived from its
thermal radio flux.

Non-thermal emission from hot stars is generally much stronger
than thermal emission, and strongly variable (the O5 supergiant 
\object{Cyg~OB2 No.~9}, for instance, was observed to vary by a factor of
ten at 6 cm by Bieging et al.~\cite{BAC89}). It also has a very different 
spectral shape, with a negative spectral index, i.e. a flux that decreases 
as a function of frequency. Such a spectrum is characteristic of
synchrotron emission from relativistic electrons gyrating in a magnetic
field. Electrons can attain relativistic speeds through first order Fermi 
acceleration in shocks (Bell \cite{Bell}, Drury \cite{D83}).
Although shock-acceleration has been most frequently used to explain
synchrotron emission from hot stars, other mechanisms such as magnetic
reconnection are also capable of accelerating particles to relativistic
speeds (e.g. Litvinenko \cite{Litvinenko}).

A fundamental question regarding non-thermal radio emission from O stars is
its correlation with binarity. If the non-thermal emitter is a member of a 
binary system, the shocks needed to accelerate the electrons
can be provided by colliding stellar winds. This is the case 
for Wolf-Rayet stars, where the correlation between non-thermal radio
emission and binarity is firmly established, and binarity appears to be a 
prerequisite for non-thermal emission (Dougherty \& Williams \cite{DW2000}).
The situation is less clear for O stars. Of the 16 currently known
non-thermal radio emitting O stars, only 10 are confirmed binaries
(Table~\ref{tb:NTO}). 

A priori, there is no need for non-thermal emitters to be binaries. The
driving mechanism of hot-star winds is known to be subject to a
line-deshadowing instability (Owocki \& Rybicki \cite{OR84}, Feldmeier
\cite{F01}). Time-dependent hydrodynamical simulations including the
line-deshadowing instability (e.g. Runacres \& Owocki \cite{RO02}), show
that it produces a wealth of shocks that in principle are capable of
accelerating electrons to high speeds. We refer to these shocks as
wind-embedded shocks. Non-thermal radio emission from hot stars was first 
interpreted as synchrotron emission from electrons accelerated in such 
wind-embedded shocks by White (\cite{White85}). Substantial improvements of 
this original idea were made by Chen (\cite{ChenPhD}) and Chen \& White
(\cite{ChenWhite94}).

\begin{table}[t]
 \caption{Binary status of non-thermal radio emitting O stars. The list is 
        compiled from
        Bieging et al. (\cite{BAC89}, a),
        Persi et al. (\cite{P85}, b), Leitherer et al. (\cite{L95}, c),
        Benaglia et al. (\cite{B01}, d), Setia Gunawan et al. (\cite{SG03}, e),
        Benaglia \& Koribalski (\cite{BK04}, f) and Drake (\cite{D90},~g). 
}
\label{tb:NTO}
\begin{center}
\begin{tabular}{lllr}
        \hline
        \hline
        Name & Spectral& Binary & NT\\
            &  Type& Status& Ref.       \\
        \hline
        \object{HD 93129A}       & O2 If + ?       &BIN$^1$        & f\\
        \object{HD 93250}        & O3 V((f))     &-             & c\\
        \object{9 Sgr}           & O4 V          &-             & a\\
        \object{CD-47$\degr$4551}& O5 If         &-             & d\\
        \object{Cyg OB2 No. 9}   & O5 If         &-             & a\\
        \object{HD 150136}       & O5 IIIn(f) + ? &BIN$^2$      & d\\
        \object{HD 168112}       & O5 III        &-             & a\\
        \object{HD 15558}        & O5 III + ?     &BIN$^3$      & a\\
        \object{Cyg OB2 No. 8A}  & O6 + O5.5     &BIN$^4$      & a\\
        \object{Cyg OB2 No. 5}   & O6 f + O7 f   &BIN$^5$    & b\\
        \object{HD 124314}       & O6 V(n)((f)) + ? &BIN$^6$      & d\\
        \object{Cyg OB2-335}     & O7 V          &-             & e\\
        \object{15 Mon}          & O7 Ve         &BIN$^8$        & g\\
        \object{HD 167971}       & O7.5 Vf + O   &BIN$^7$     & a\\
        \object{$\delta$ Ori A}  & O9.5 II + ?    &BIN$^9$        & g\\
        \object{$\sigma$ Ori AB} & O9.5 V + ?    &BIN$^{10}$     & g \\
        \hline
\end{tabular}
\end{center}
$^1$ Walborn (\cite{W02}); $^2$ Garmany et al. (\cite{GCM80}); $^3$ Garmany \& Massey
(\cite{GM81}); $^4$ De Becker et al. (\cite{DB04b}); $^5$ Torres-Dodgen et al. (\cite{TD91}); 
$^6$ Gies (\cite{G87}); $^7$ Leitherer et al. (\cite{L87});
$^8$ Gies et al.~(\cite{G93}); $^9$ Harvey et al.~(\cite{H87}); $^{10}$~Bolton~(\cite{B74})
\end{table}

This is the third in a series of papers investigating the non-thermal radio
emission from single stars. The general aim is to determine the
properties that wind-embedded shock models must have to reproduce the
observed non-thermal radio emission. By this means, we are also able to
shed some light on the fundamental question of binarity, as it will become 
clear to what extent the wind-embedded shock model is able to explain the 
observations. Such an indirect approach is necessary because it is difficult, 
in general, to confirm observationally that a star is single. A failure of the 
wind-embedded shock model to explain the observations would strongly suggest 
that colliding winds are required to produce non-thermal radio emission.

In Van Loo et al. (\cite{PaperI}, hereafter Paper~I), we introduced a simple 
parametrised model. The number distribution of synchrotron-emitting electrons 
was given by a power law, both as a function of momentum and distance. 
The emission region was assumed to extend {\em continuously} out to an
outer radius $R_{\rm max}$. This assumption is open to criticism, 
because cooling mechanisms such as inverse Compton scattering can limit the 
emission to potentially thin layers behind the shocks (Chen \cite{ChenPhD}). 
This effect was included in Van Loo et al.~(\cite{PaperII}, hereafter Paper~II). 
In that paper we found that the emitting layers can indeed be quite narrow and 
are also few in number, because the strongest shocks tend to dominate the emission. 
In the present paper, we further extend the model to take into account the fact 
that shocks gradually weaken as they move out with the flow. Specifically, we 
use results of time-dependent hydrodynamical simulations as input to our 
synchrotron-emission models.

The comparison of synchrotron spectra with hydrodynamical simulations of 
instability-generated structure has the additional advantage of providing 
supplementary observational input for such simulations. 
Although the theoretical argument for the line-deshadowing instability is extremely strong 
(Owocki \cite{O91}), and the 
instability-generated structure is still the best explanation for, e.g., 
soft X-ray emission, black troughs, and clumping inferred from density-squared
diagnostics (e.g. Owocki \cite{O91}),
the small-scale inhomogeneities it generates are not easy 
to detect. Therefore, any additional observational information 
about such structure would be most welcome.

The remainder of this paper is organised as follows. In Sect.~\ref{sect:model} 
we introduce the model. In Sect.~\ref{sect:results} we conclude that it fails 
to explain the \object{Cyg OB2 No. 9} observations. We explore various 
``escape routes" in Sect.~\ref{sect:discussion}. Conclusions are given in 
Sect.~\ref{sect:conclusions}.

\section{The model}\label{sect:model}
\subsection{Layered emission}
As the present model is based on the one from Paper~II, we briefly review 
the properties of this latter model. The synchrotron emission is produced by
electrons accelerated in wind-embedded shocks, produced by the line-deshadowing 
instability in a single star wind. The acceleration is not calculated ab
initio, but a power law is assumed for the  momentum distribution of
electrons. 

In principle, the pressure from the relativistic particles can 
influence the structure of the shocks, and can even lead to their destruction
(Drury \cite{D83}).
Incorporating this influence in time-dependent hydrodynamical simulations is
beyond the scope of the present work. We therefore neglect the effect
of relativistic particles on the hydrodynamics (this is referred to as the
``test-particle approach"). The result of incorporating this influence in 
time-dependent hydrodynamical simulations would be to weaken the shocks, 
thereby making it more difficult to produce the extended regions of non-thermal 
emission needed to explain the observations. As we shall find, it is already 
difficult to produce
the observed emission based on present shock strengths 
(Sect.~\ref{sect:hydro input}), so our conclusions will be all the more 
valid if the feedback of relativistic particles on the hydrodynamics is 
included.

We adopt the magnetic field expression of Weber \& Davis (\cite{WD67}).
This means that, at large distances from the star, the magnetic field
is almost perpendicular to the shock normal. Jokipii (\cite{J87}) showed
that this configuration (i.e. quasi-perpendicular shocks) is the most
efficient for particle acceleration. Details are given in Van Loo
(\cite{SVLPhD}).

The electrons are subject to a number of cooling mechanisms, most notably inverse 
Compton scattering and adiabatic expansion. Cooling has an important effect both 
at the shock front, and downstream. At the shock front, the maximum attainable 
energy is set by the balance between acceleration and cooling, imposing a 
high-momentum cut-off on the distribution. Downstream from the shock, the acceleration 
ceases and cooling rapidly removes the energy of relativistic electrons.
The emission is then expected to come not from a continuous volume, but from a 
(possibly thin) layer behind each shock. Cooling also modifies the shape of the 
momentum distribution close to the cut-off. Away from the cut-off it retains its 
power-law shape.

In Paper~II, we also showed that the strongest shocks dominate the
emission. This means that the number of synchrotron emission layers that
contribute substantially to the observed emission is small. It is therefore natural 
to consider first the case of a single shock. A single-shock model is 
characterised by four free parameters: the position $R_{\rm S}$ of the shock, 
its velocity jump $\Delta u$ and its compression ratio $\chi$, and the surface 
magnetic field $B_*$. The fluxes include a small but significant thermal contribution 
(25\% of the total flux at 2~cm). For a complete discussion of the model, we refer 
to Paper~II.

To describe the distribution behind the shock front, we need to know where
the electrons were accelerated and how they are cooled. The relativistic electrons 
move away from the shock front with the gas. In terms of the velocity jump 
$\Delta u$ and the compression ratio $\chi$, the post-shock speed in the frame of 
the shock can be expressed as
\begin{equation}\label{eq:outflow}
	u_2=\frac{\Delta u}{\chi-1},
\end{equation}
which follows directly from the Rankine-Hugoniot relations. As the gas flows
a distance $\Delta R_{\rm S}$ away from the shock at a speed $u_2$, the shock has 
travelled a much larger distance $\approx \Delta R_{\rm S} v_{\infty}/u_2 $, 
where $v_{\infty}$ is the terminal speed. As $v_{\infty}/u_2 \sim 100$,
the electron distribution in a fairly narrow layer behind an outer-wind 
shock reflects wind conditions over a wide range in distance. The outer electrons 
of a typical synchrotron-emitting layer (4~$R_*$ wide, behind a shock at $500~R_*$), 
will have been accelerated at a distance $\sim 100~R_*$ from the central star.
Of course, they are subjected to severe cooling during the time-interval needed 
to reach the outer wind.

In Paper~II, we assumed a constant shock velocity jump and compression ratio, mainly 
to keep the number of free parameters in the model as small as possible. However, 
hydrodynamical simulations show that both the shock velocity jump and the compression 
ratio decline with radius (see Fig. \ref{fig:hydro:chi}). The shock velocity jump 
appears as a strong scaling factor in this model, with the flux proportional to 
$\Delta u^3$. The dependence of the synchrotron flux on the compression ratio 
$\chi$ is somewhat more complicated, but is also important (Van Loo \cite{SVLPhD}). 
Generally, the flux increases rapidly with $\chi$. In view of the wide range in 
radius reflected by the electron distribution, one can therefore expect that introducing 
the radial decline of the velocity jump and the compression ratio has important 
consequences for the models.

Due to the free-free opacity of the dense stellar wind, any radio emission 
(thermal or non-thermal) that is emitted too close to the star will be absorbed.
Roughly speaking, the emergent radio flux originates beyond the radius of unit 
free-free optical depth. This radius can be viewed as a kind of ``radio photosphere" 
of the stellar wind. Due to the wavelength-squared dependence of the free-free 
opacity, the size of the radio photosphere increases with wavelength.
For example, in the well-studied non-thermal radio emitting O star 
\object{Cyg OB2 No. 9}, the radio photosphere is $90~R_*$ at 2~cm (15~GHz) and 
$195~R_*$ at 6~cm (5~GHz). The emergent flux of a non-thermal emitter 
is determined by the tight interplay between non-thermal emission and thermal
absorption. This is what gives the models their great diagnostic value.
A more detailed discussion of the radio formation region is given in
Paper~I.

\subsection{Hydrodynamical input}\label{sect:hydro input}
Since the pioneering models of Owocki et al. (\cite{OCR}), many hydrodynamical 
simulations of hot-star winds have included the line-deshadowing instability 
(e.g. Feldmeier \cite{F95}, Dessart \& Owocki \cite{DO02}, Runacres \& Owocki 
\cite{RO02}). These simulations show that within a few stellar radii of the 
surface, the flow becomes highly structured, with gas concentrated in dense 
clumps, and pervaded by strong shocks. These simulations are restricted to 
distances smaller than $100~R_*$, which is not far enough to provide input for 
synchrotron-emission models. For simulations extending out to large radii,
Runacres \& Owocki (\cite{RO04}) introduced a periodic-box formalism for 
the outer evolution of structure in hot-star winds, which enabled them
to follow structure out to distances of more than 1000~$R_*$.
\begin{figure}
\resizebox{\hsize}{!}{\includegraphics{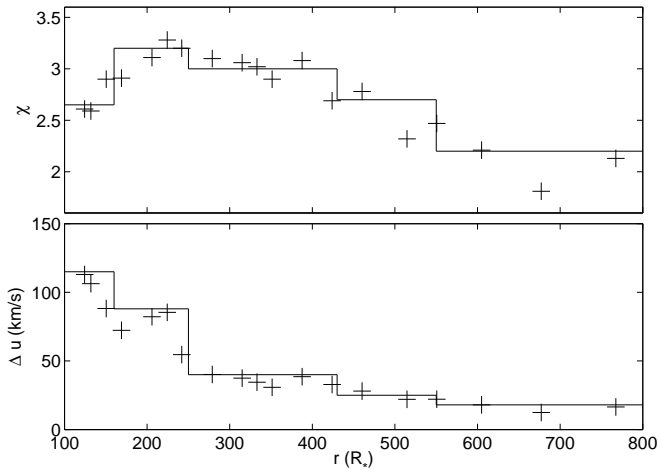}}
\caption{
     Compression ratio (plus signs, upper panel) and velocity jump (plus signs, 
     lower panel) of a typical shock in the periodic-box model of Runacres \& Owocki 
     (\cite{RO04}), as a function of radius. The solid lines represent the step 
     function approximations used in the code. 
}
\label{fig:hydro:chi}
\end{figure}

In Fig.~\ref{fig:hydro:chi} we show the velocity jump and compression
ratio of a typical strong shock in a periodic-box model as it moves out with
the gas. The shock velocity jump steadily decreases from $\ga 100~{\rm km\,s^{-1}}$ 
at 100~$R_*$ to $\approx 55~{\rm km\,s^{-1}}$ at 250~$R_*$ and further to
$\approx 20~{\rm km\,s^{-1}}$ beyond 500~$R_*$.  The compression ratio rises 
from $\approx 2.6$ at 100~$R_*$ to 3.3 at 250~$R_*$, and then decreases to values 
$\approx 2.2$ beyond 500~$R_*$. The compression ratio increases 
between 100 and 200~$R_*$ because the strong forward shock that we follow overtakes 
a few weak forward shocks and merges with them. 

As it takes some time to cool relativistic electrons below radio-emitting energies,
the inner-wind accelerated electrons can contribute significantly to the radio 
emission.  Furthermore, since the emissivity rapidly declines with radius due to the
decreasing shock velocity jump and compression ratio, the emergent flux is weighted 
in favour of electrons accelerated in the inner wind, and the radial dependence of the 
velocity jump and the compression ratio must be included in the model. For the sake of 
simplicity, we approximate the radial dependence of $\Delta u$ and $\chi$ by step 
functions of the form shown in Fig.~\ref{fig:hydro:chi}.

\section{Results}\label{sect:results}
\subsection{Single shock}\label{subsect:single shock}
In a single-shock model, 
explicitly taking into account the radial dependence of $\Delta u$ and $\chi$ has
an important effect on the emergent flux.  In Paper~II we showed that a single shock located at 
$\sim 600~R_*$ 
can explain the observed non-thermal emission of \object{Cyg OB2 No. 9}, 
 albeit with a much larger 
velocity jump ($290~{\rm km\,s^{-1}}$) than predicted by hydrodynamical simulations. From 
Fig.~\ref{fig:hydro:chi} we derive that a typical shock at that distance has a compression ratio
$\chi = 2.2$ and a velocity jump $\Delta u = 18~{\rm km\,s^{-1}}$. In Paper~II we assumed that 
the shock strength is the same everywhere in the wind, whereas the hydrodynamical simulations 
predict a weakening of the shock. It is to be expected, for a weak outer-wind shock, that its 
past history as a stronger inner-wind shock affects the energy distribution of the accelerated 
electrons behind it.

\begin{figure} 
\resizebox{\hsize}{!}{\includegraphics{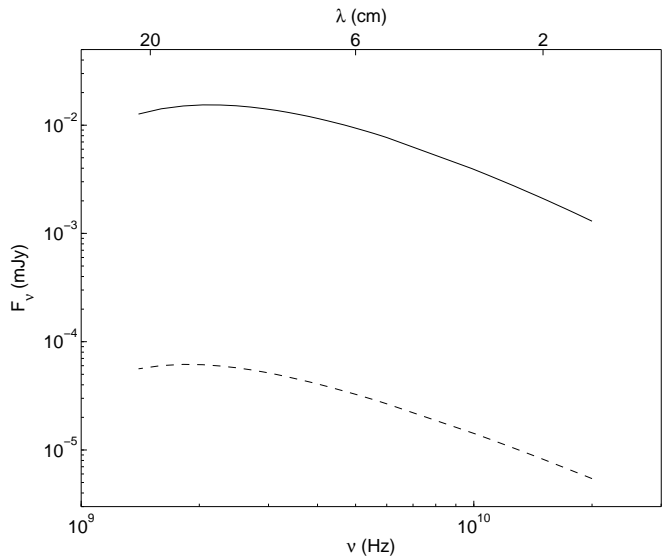}}
\caption{
     Pure synchrotron flux for a single-shock model with declining
     $\Delta u$ and $\chi$ given by the step functions in Fig.~\ref{fig:hydro:chi} 
     (solid line), compared to a model with constant shock strength $\chi = 2.2$ and
     $\Delta u = 18~{\rm km\,s^{-1}}$ (dashed line), i.e. a model where the history of 
     the shock has not been taken into account. The shock is located at 
     $R_{\rm S} = 600~R_*$. The surface magnetic field is $B_* = 200$~G. Stellar
     and wind parameters are as in Paper~II.
}
\label{fig:single:IvsII}
\end{figure}

In  Fig.~\ref{fig:single:IvsII} we compare
the pure synchrotron flux (without the thermal contribution) generated by
such a single shock assuming a constant shock strength throughout the wind (dashed line), 
and  the  flux emitted by the same shock when its history is taken into account (solid line).
This last model produces significantly more emission (by a factor of $\sim$ 200). This is 
because, when the shock was closer to the star and consequently stronger, it 
produced a population of electrons that are still sufficiently energetic, in spite 
of strong inverse-Compton cooling, to produce synchrotron emission. 

When the history of the shock is taken into account, a single shock can still explain the 
observed non-thermal spectrum of \object{Cyg~OB2 No.~9}. To 
fit the observed flux, we need to adjust the magnitude of the shock velocity jump. 
We do this by defining the velocity jump at radius $r$ as 
\begin{equation}\label{eq:f}
    \Delta u(r) = \Delta u_{\rm max} f(r), 
\end{equation}
where $f(r)$ is the velocity jump from Fig.~\ref{fig:hydro:chi}, divided by its highest 
value (i.e. $f(r) \le 1$), and $\Delta u_{\rm max}$ is a scaling factor that sets 
the magnitude of the velocity jump.  

\begin{figure}
\resizebox{\hsize}{!}{\includegraphics{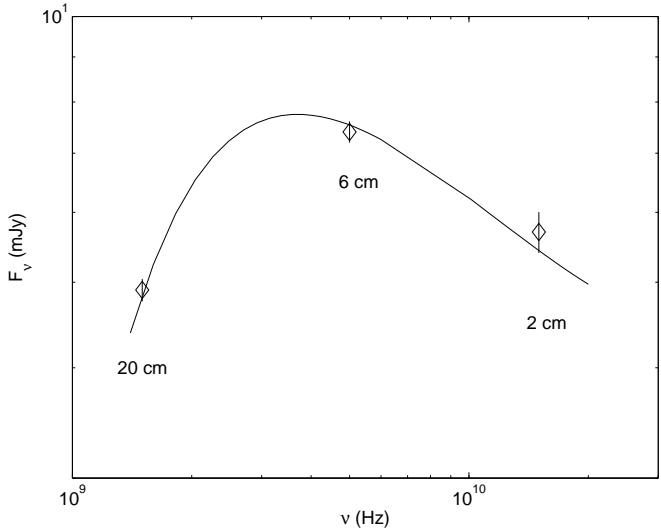}}
\caption{
     A single-shock model (solid line) that fits the observations
     (diamonds) of Cyg~OB2 No.~9. The vertical stripes are the error bars.
     The shock velocity jump is $\Delta u(r) = \Delta u_{\rm max} f(r)$, 
     as defined in Eq.~(\ref{eq:f}). The fit parameters are 
     $\Delta u_{\rm max} = 1050~{\rm km\,s^{-1}}$, $R_{\rm S}=602~R_*$, 
     $\chi =4$ and $B_* =200$~G.
}
\label{fig:single shock}
\end{figure}

In Fig.~\ref{fig:single shock} we show a single-shock model that fits the observed
spectrum. For simplicity, the compression ratio $\chi$ is taken to be 4 at all
radii. The value of $\Delta u_{\rm max}$ is $1050~{\rm km\,s^{-1}}$, corresponding to
a local velocity jump at $R_{\rm S}=602~R_*$ of $164~{\rm km\,s^{-1}}$. This is lower 
than the value determined in Paper~II (due to the significant amount of inner wind 
accelerated electrons in the emission layer) but still an order of magnitude higher than 
predicted by hydrodynamical simulations.  With a decreasing $\chi$, the value obtained 
for $\Delta u$ would be even higher. The only way to fit the observations with smaller 
velocity jumps is to consider multiple shocks.

\subsection{Multiple shocks}
\subsubsection{Constant $\chi$ and $\Delta u$}
\label{sect: constant strength}
The synchrotron emission in our model is proportional to the third power of the shock 
velocity jump (Paper~II). This allows us to estimate the flux of a multiple-shock 
model with a constant compression ratio and velocity jump from a single-shock model.
The condition for a collection of $N$ identical shocks with velocity jump $\Delta u$, 
distributed throughout the wind, to produce the same flux as a single shock with velocity 
jump $\Delta u_{\rm single}$ is simply expressed as
\begin{equation}\label{eq:multiple}
    N \Delta u^3 = \Delta u_{\rm single}^3.
\end{equation}
This offers a way of lowering the velocity jump needed to explain the observations: 
one just adds shocks until the velocity jump is in the range of the hydrodynamical 
predictions. In Paper~II, we fitted the observations by a single shock with 
$\Delta u_{\rm single} = 290~{\rm km\,s^{-1}}$. Using the above equation we estimated 
that 200 shocks with $\Delta u = 50~{\rm km\,s^{-1}}$ should produce the same
flux.

When the shock velocity jump and/or compression ratio decrease as a function of radius,
as predicted by hydrodynamical simulations, Eq.~(\ref{eq:multiple}) can no longer be 
applied.  In that case, we find that our model cannot reproduce the observations
(Sect.~\ref{subsubsect:declining strength}). To understand why this is so, it is 
useful to consider first the {\em  cumulative} flux for a model with a {\em constant} 
shock velocity jump and compression ratio. (The cumulative flux at a position $r$ in 
the wind is defined as the emergent flux due to all the shocks between the stellar
surface and $r$.)
\begin{figure}
\resizebox{\hsize}{!}{\includegraphics{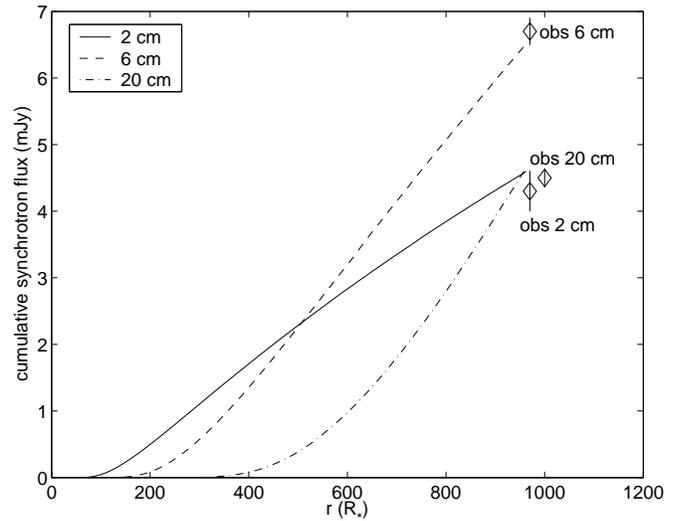}}
\caption{
    Cumulative fluxes for a model with constant compression ratio and velocity
    jump, at 2 cm (solid line), 6 cm (dashed) and 20 cm (dashed-dotted). The 
    diamonds are the observed synchrotron fluxes and also the error bar on the 
    observations is given. The model parameters are $\Delta u=71~{\rm km\,s^{-1}}$, 
    $\chi=3.7$ and $B_*=200$~G ($\Delta u$ and $\chi$ are constant in the wind). 
    A shock is present every $12~R_*$.
}
\label{fig:constant strength}
\end{figure}
In Fig.~\ref{fig:constant strength} we show the cumulative fluxes at different radio
wavelengths for an equidistant distribution of shocks, with a shock at every 12~$R_*$
(The influence of the adopted shock spacing is discussed in Sect.~\ref{sect: spacing}).
All shocks have $\chi=3.7$ and $\Delta u=71~{\rm km\,s^{-1}}$. The surface magnetic 
field is $B_* = 200$~G. Only the synchrotron flux is shown. The ``observed"  synchrotron 
flux  is obtained by subtracting the theoretical thermal flux (using the Wright \& Barlow
formalism) from the observed radio flux. The cumulative synchrotron flux at 2~cm (solid 
line) is zero for distances below $\sim 50~R_*$, where all synchrotron emission is 
absorbed by the free-free opacity of the wind, and then rises sharply. The flux at 
6~cm (dashed) begins to rise further out, due to the larger free-free optical depth. For 
the same reason, the 20~cm begins to rise even later.

The cumulative flux increases more steeply at larger wavelengths. The reason is twofold. 
First, there is the intrinsic spectrum of synchrotron emission: the flux at larger 
wavelengths is produced by less energetic electrons, of which there are more by virtue
of the shape of the momentum distribution. Second, inverse-Compton cooling is 
less severe at larger wavelengths, so that the synchrotron-emitting layer is wider.

\subsubsection{Declining $\chi$ and $\Delta u$}
\label{subsubsect:declining strength}
Let us now see what happens when the shock velocity jump and compression ratio decrease 
as a function of radius, as predicted by hydrodynamical simulations.  We can no longer 
use Eq.~(\ref{eq:multiple}) to determine how many shocks are needed, but plotting the 
cumulative flux allows us to judge when we have a good fit to the observations, without 
requiring the number of shocks to be pre-specified.

\begin{figure}
\resizebox{\hsize}{!}{\includegraphics{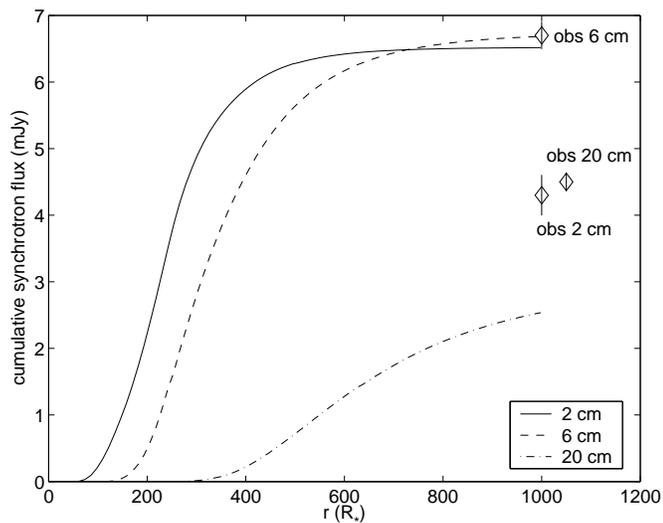}}
\caption{
     Cumulative fluxes for a model with decreasing compression ratio and velocity
     jump, at 2 cm (solid line), 6 cm (dashed) and 20 cm (dash-dotted). The diamonds 
     are the observed synchrotron fluxes and also the error bar on the observations 
     is given. The model parameters are $B_*=200$~G, 
     $\Delta u_{\rm max}=194~{\rm km\,s^{-1}}$ and $\chi$ is taken from 
     Fig.~\ref{fig:hydro:chi}. A shock is present every $12~R_*$.
}
\label{fig:refmodel}
\end{figure}

In Fig.~\ref{fig:refmodel} we show the cumulative flux for a model that
is similar to the model shown in Fig.~\ref{fig:constant strength}, except for the 
decreasing shock strength. The compression ratio of the shocks is given by
Fig.~\ref{fig:hydro:chi} (top panel). The velocity jump is given by the lower panel of 
the same figure, but a scaling factor $\Delta u_{\rm max}$ is introduced, as in 
Sect.~\ref{subsect:single shock}, to fit the observed flux at 6~cm. In the model 
shown, $\Delta u_{\rm max} = 194~{\rm km\,s^{-1}}$, which represents a $70 \%$ 
increase with respect to the hydrodynamical predictions. The cumulative fluxes are zero 
below the radio photosphere, and then rise steeply. Contrary to the model with constant 
velocity jump and compression ratio, there is no discernible difference in the rate
of increase of the fluxes at 2 and 6 cm. If the shock strength were constant,
the 6~cm flux would rise faster than the 2~cm flux (see Sect.~\ref{sect: constant strength}).
This is now compensated by the fact that the 2 cm flux is formed at smaller radii, where 
the shocks are stronger. 

Both the 2 and 6 cm fluxes level off at large radii, because the compression ratio 
and velocity jump become so small that the emitted synchrotron radiation is negligible. 
The fluxes at 2 and 6 cm reach approximately the same asymptotic value, resulting in 
a flat synchrotron spectrum. After adding the thermal contribution, the emergent 
spectrum has a positive slope, contrary to observations. At 20 cm, the model 
significantly underestimates the observation. Extending the model to larger radii 
would increase the 20~cm flux somewhat, but would have no effect on the 2 and 6~cm flux.

In summary, we find that a model where the velocity jump and compression ratio are 
taken from hydrodynamical simulations does not reproduce the observed radio spectrum,
because the strong shocks close to the star contribute more to the flux at shorter 
wavelengths (where the free-free opacity is smaller) than at longer wavelengths. 
Any model where the emissivity is a strongly decreasing function of radius will 
therefore have a tendency toward a positive spectral index. Conversely, to reproduce 
the observed negative spectral index, any model should have an emissivity that is a 
weakly decreasing function of radius.

\section{Discussion}\label{sect:discussion}
The hydrodynamical simulations from which we take the shock velocity jump and 
compression ratio are subject to considerable uncertainty. Foremost, they are 
performed in one dimension, whereas the stellar wind is expected to contain a 
fair amount of lateral structure (e.g. due to Rayleigh-Taylor instabilities, 
see Dessart \& Owocki \cite{DO03}). Moreover, no attempt has been made to model 
a particular star. The results of these simulations should therefore not be taken 
at face value. To further quantify the discrepancy between the hydrodynamical 
predictions and the radio observations, let us see by which modifications the two 
may be reconciled or, in other words, which ``escape routes" may lead us away from 
the discrepancy, and whether any of these escape routes is physically plausible.

\subsection{Shock spacing and outer boundary of the synchrotron-emitting
region}
\label{sect: spacing}
In Sect.~\ref{sect:results} we adopted a shock spacing of $12~R_*$. This
value is somewhat arbitrary but is based on the spacing found in
hydrodynamical simulations, and on the fact that only the strongest shocks
will contribute substantially to the synchrotron emission. Reducing the
spacing from $12~R_*$ to $4~R_*$, which is about the smallest distance 
between two adjacent outer wind shocks in the hydrodynamical simulations,
does not change the slope of the emergent synchrotron spectrum. The only
effect of the closer spacing is a lower value of $\Delta u_{\rm max}$, due
to the larger number of shocks in the wind.

To produce the observed negative spectral index, the radial decrease of the synchrotron 
emission should be weak. Such a slow decrease can be obtained by a closer shock spacing 
in the outer wind than in the inner wind (to counteract the decrease of the shock 
strength). Physically this corresponds to a slowing down of the shocks as they move out 
with the wind. Such a slowing down can indeed be expected from momentum conservation. 
However, the associated change in velocity is minimal. Both the shell and the material 
it sweeps up are moving at roughly the terminal speed, with velocity differences of the 
order of  a few times the sound speed. We find that, to reproduce the observed spectral 
index, the spacing of the shocks in the 20 cm formation region should be less than half 
of that  in the 2 cm formation region. This corresponds to a decrease in shock speed 
(measured in the stellar rest frame) of a factor of two or more, which is not predicted 
either by detailed hydrodynamical simulations or by the momentum conservation argument 
above.

The adopted outer boundary of the synchrotron-emitting region could in principle influence 
the spectral index predicted by the model.  However,
the fact that the cumulative flux in Fig.~\ref{fig:refmodel} levels off at 
large radii shows that the spectral shape cannot be changed  by extending the 
synchrotron-emitting region to larger radii (as the contribution at large radii becomes 
negligible).

We conclude that our model cannot be made to produce a negative spectral
index by increasing the number of shocks in the wind, either by changing the
shock spacing or  by adopting a larger outer boundary of the synchrotron-emitting region.

\subsection{Rate of decline of $\chi$ and $\Delta u$} 
To reproduce the observed spectra, any model should have an emissivity that
does not decline too rapidly as a function of radius. As the emissivity is
directly related to the compression ratio and velocity jump of the shocks
producing the emission, making these properties a weaker function of radius
is a natural way of obtaining a more slowly declining emissivity.
As it is difficult to disentangle the effects of a declining velocity jump
and a declining compression ratio, we adopt a constant compression ratio and 
see how weak a function of radius the velocity jump should be to produce a
\begin{figure}
\resizebox{\hsize}{!}{\includegraphics{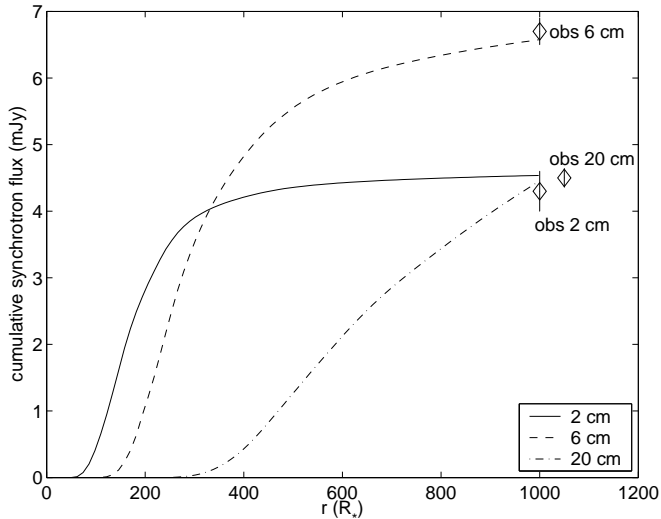}}
\caption{
     Cumulative fluxes for a model with a ``flattened" $\Delta u (r)$,
     at 2 cm (solid line), 6 cm (dashed) and 20 cm (dash-dotted). The diamonds 
     are the observed synchrotron fluxes and also the error bar on the observations 
     is given. The ratio between the maximum and minimum $\Delta u$ is 3. The other 
     parameters are $\Delta u_{\rm max}=835~{\rm km\,s^{-1}}$, $\chi=2$ and $B_*=200$~G.
     A shock is present every $12~R_*$.
}
\label{fig:escape:inout}
\end{figure}
negative spectral index. Therefore, we make $\Delta u (r)$ a weaker function of 
$r$, by replacing $f(r)$ in Eq.~(\ref{eq:f}) with $\tilde f(r)$, defined as
\begin{equation}\label{eq:f flat}
     \tilde f(r) = 1- \epsilon \left[ 1- f(r)\right] \hspace*{3cm} 
     0 < \epsilon < 1
\end{equation}
where $\epsilon=1$ corresponds to the original $f(r)$ and $\epsilon=0$
to a constant velocity jump. In Fig.~\ref{fig:escape:inout} we show the 
cumulative flux for a model where $\Delta u (r)$ has been ``flattened" 
in this manner and a constant compression ratio $\chi=2$. The ratio between 
the maximum and minimum values of $\Delta u$ is 3 (corresponding to 
$\epsilon=0.79$), as opposed to 6.4 in Fig.~\ref{fig:hydro:chi}. This model
explains the observations at all wavelengths, although the required 
$\Delta u_{\rm max}$ is much higher than the hydrodynamical predictions. 
The observations can also be reproduced with a higher compression ratio $\chi=3$
and an even flatter $\Delta u (r)$, with a ratio of 1.5 between its maximum 
and minimum.

In these models, we assumed a constant compression ratio, whereas the
hydrodynamical simulations predict a decline (Fig.~\ref{fig:hydro:chi}).
With a declining compression ratio, $\Delta u (r)$ must be even 
flatter.

Indeed, we find that, with $\chi(r)$ as defined in Fig.~\ref{fig:hydro:chi},
$\Delta u$ must be constant to fit the observations. If $\chi (r)$ is 
flattened in the same way as was done above for $\Delta u (r)$
(Eq.~(\ref{eq:f flat})), a fit can be obtained if the ratio between maximum 
and minimum is 1.3 or less for both $\Delta u$ and $\chi$. 

The upshot of the above is that the radio spectrum of \object{Cyg~OB2 No.~9}
can only be explained with our model with a modest ($\le 30\%$) decrease of the
velocity jump and compression ratio. The range in radius contributing to the
emergent flux at 2, 6 and 20~cm is $\sim 30-1\,000~R_*$. Although not absolutely 
impossible, it is unlikely that shocks would have such a small decrease in 
velocity jump and compression ratio over such a large distance.

\subsection{Re-acceleration}
Our results show that it is difficult for a wind-embedded shock model to
reproduce an observed non-thermal radio spectrum, if the radial decline 
of the shock strength is in agreement with hydrodynamical predictions. 
This may seem at odds with some previous models that successfully reproduce 
the observed non-thermal spectra. The most advanced of these models is
given in Chen (\cite{ChenPhD}) and Chen \& White (\cite{ChenWhite94}). 
Their model can successfully reproduce the observed non-thermal spectrum of 
both \object{9 Sgr} and \object{Cyg OB2 No. 9}. It includes adiabatic and 
inverse-Compton cooling.  As a result, electrons cannot travel very far from 
the acceleration site, and the electrons have to be accelerated {\em in situ} 
in the radio-emitting region. The emission is then also confined to layers 
behind the shocks. In that respect, the  Chen \& White (\cite{ChenWhite94}) 
model is very similar to the model presented in the present paper.

The fundamental difference between Chen \& White and the present model 
is re-acceleration of electrons. If a fluid element goes through a number
of successive shocks as it moves outward, then relativistic electrons
in the outer wind will, on average, have passed through more shocks than
inner-wind electrons. Consequently, these outer-wind electrons are more
energetic than they would have been at the same position in the wind,
had they passed through only one shock. Therefore, re-acceleration
substantially slows down the outward decrease of the synchrotron
emissivity and is expected to facilitate a negative spectral index.
Indeed, with a simplified model of re-acceleration, where we artificially
slow down the radial decrease of the relativistic electron density, we
easily reproduce the observations.

Whether or not re-acceleration occurs depends on the adopted hydrodynamical 
model. In Chen \& White (\cite{ChenWhite94}), the hydrodynamical model is from
Lucy (\cite{Lucy}), supplemented with a snowplough model to describe structure in 
the outer wind (Chen \cite{ChenPhD}).  In Lucy (\cite{Lucy}), all shocks are 
forward shocks. As was pointed out by White (White \& Chen \cite{WhiteChen94}, 
discussion), a fluid element in the Lucy model is accelerated to the terminal 
speed only in the shock jumps. Therefore, any given fluid element goes through 
a large number of shocks.  Relativistic electrons are then accelerated at one 
shock front, cool as they travel downstream to the next shock, and are re-accelerated
there. A population of new, freshly accelerated relativistic electrons is also 
created at each shock.

Lucy's model is entirely phenomenological, however. When the line-deshadowing 
instability is introduced in a model that solves the time-dependent equations 
of hydrodynamics, the physical picture of wind structure is quite different. 
The fundamental consequence of the instability is the omnipresence of strong 
rarefaction waves of fast, rarefied gas. These rarefactions are terminated by 
a reverse shock (much as when gas runs into a solid wall), forming the inner 
boundary of a dense shell. As this dense shell is impacted by faster shells, 
it develops a forward shock on the outside. The result is that shocks occur in 
pairs at either side of a dense shell. Gas is fed through the shocks into the shell
and then remains trapped there. A fluid element therefore only encounters a 
single shock, making re-acceleration impossible.

The simulations on which the above analysis is based are one-dimensional. In 
a three-dimensional situation the trapping of the gas is less absolute, as gas 
can in principle escape sideways from a shell fragment or blob and encounter 
another shock. It is hard to make a quantitative estimate of the importance of 
such an effect. As the radiative driving and the associated instability are 
strongest in the radial direction, one can expect the azimuthal velocity to be 
relatively small, even though instabilities such as Rayleigh-Taylor are expected 
to disrupt the lateral coherence of the structure. One situation where the 
azimuthal velocity could  become more important is the case of rarefied gas 
flowing around a dense blob. However, this gas has a low density and should 
therefore not contribute much to the emission. It therefore seems likely that 
most of the dense gas will see few shocks, which impairs the main mechanism 
by which Chen \& White (\cite{ChenWhite94}) were able to produce a negative 
spectral index.

\subsection{Smaller radio photosphere}
\begin{figure}
\resizebox{\hsize}{!}{\includegraphics{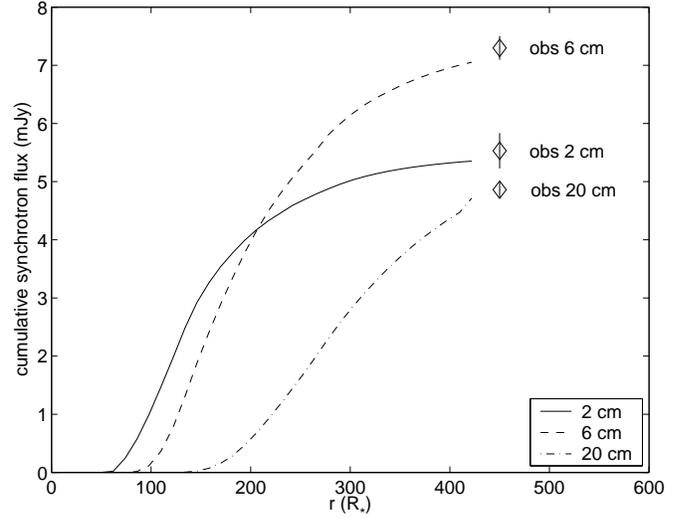}}
\caption{
    Cumulative fluxes for a synchrotron model with an artificially low 
    mass-loss rate, at 2 cm (solid line), 6 cm (dashed) and 20 cm (dash-dotted). 
    The diamonds are the observed synchrotron fluxes and also the error bar on 
    the observations is given. The model parameters are 
    $\dot{M}=4.0\times10^{-6}~{\rm M_{\sun}\,yr^{-1}}$,
    $\Delta u_{\rm max}=152~{\rm km\,s^{-1}}$, $\chi=3$ and $B_*=200$~G. We 
    extrapolated $\Delta u$ inward, so that, for $r<30~R_*$, 
    $\Delta u=1.74~\Delta u_{\rm max}$ and, for $30~R_*<r<100~R_*$, 
    $\Delta u=1.39~\Delta u_{\rm max}$. A shock is present every $12~R_*$.
}
\label{fig:escape:Mdot}
\end{figure}

The main impediment for a strongly decreasing synchrotron emissivity to produce 
a negative spectral index is the difference in the radio photosphere at 2 and 
6~cm, which causes the 2 cm flux to benefit from strong emission that is 
inaccessible to the 6~cm flux. This difference in radio photosphere increases 
with density, and hence with the mass-loss rate. It can therefore be expected 
that a lower mass loss rate may produce a negative spectral index even with a 
strongly declining synchrotron emissivity. In Fig.~\ref{fig:escape:Mdot} we show 
the cumulative flux for 
$\dot{M} = 4.0 \times 10^{-6}~{\rm M_{\sun}\,yr^{-1}}$, which is a 
factor 5 lower than the observed $H\alpha$ mass-loss rate we adopted in the 
rest of this paper. By restricting the outer boundary of the synchrotron-emitting 
region to $430~R_*$, we find that the model reproduces the observed radio spectrum.

There is no reason to assume, however, that the mass-loss rate of 
\object{Cyg~OB2 No.~9} can simply be lowered by a factor of five. Although the
mass-loss rate is extremely high for an O star, \object{Cyg~OB2 No.~9} is
also extremely bright ($L_*=1.74\times10^6~{\rm L_{\sun}}$), and it obeys 
the normal wind-momentum luminosity relation for O stars. It is widely accepted 
that observed O-star mass-loss rates might be overestimated by a factor of a few 
due to clumping (e.g. Massa et al.~\cite{M03}, Hillier et al.~\cite{H03}). The 
hydrodynamical simulations on which we have based our model obviously predict a 
large amount of clumping. However, it is important to realise that a reduced
mass-loss rate due to optically thin clumps does not reduce the free-free optical depth
{\em if the clumped model is to produce the same thermal flux} 
(see Appendix~\ref{sect:effect clumping}). In other words, assuming a clumped 
wind does not provide a way to reconcile the hydrodynamical predictions with the 
radio observations.  Other mechanisms such as porosity (e.g. Owocki et al.
\cite{OGS04}), where wind material is packed together also in the lateral 
direction,  may play a r\^ole when individual clumps become optically thick.
Owocki et al. (\cite{OGS04}) show that for optically thick clumps the opacity 
is effectively reduced. Porosity is characterised by an additional parameter 
(the clump size or equivalent) which complicates the relation between the clump 
properties and the radio flux. It is as yet unclear to what extent hot-star winds 
are affected by porosity, if at all.

The O4 main-sequence star \object{9 Sgr} has a much lower mass-loss rate
($\dot{M} = 2.4 \times 10^{-6}~{\rm M_{\sun}\,yr^{-1}}$) than
\object{Cyg OB2 No. 9}. Therefore, one could expect its spectrum to be
compatible with the predicted decline of the shock strength.
\begin{figure}
\resizebox{\hsize}{!}{\includegraphics{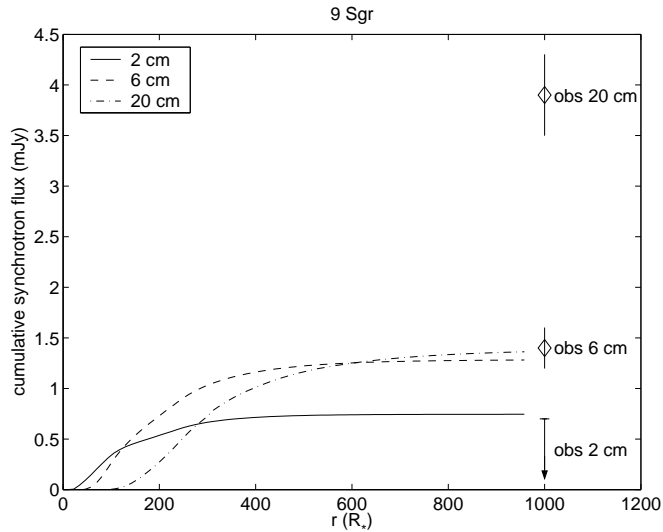}}
\caption{
     Cumulative fluxes at 2 cm (solid line), 6 cm (dashed)
     and 20 cm (dash-dotted) for \object{9 Sgr}. The diamonds are the observed 
     synchrotron fluxes and also the error bar on the observations is given. 
     The arrow denotes the upper limit at 2 cm.  The model parameters are
     $\dot{M}=2.4\times10^{-6}~{\rm M_{\sun}\,yr^{-1}}$, 
     $\Delta u_{\rm max}= 130~{\rm km\,s^{-1}}$, $\chi=3$ and $B_*=200$~G.
     We extrapolated $\Delta u$ inward in the same way as for 
     Fig.~\ref{fig:escape:Mdot}. A shock is present every $12~R_*$.
}
\label{fig:9 Sgr}
\end{figure}
Figure~\ref{fig:9 Sgr} shows the cumulative flux at 2, 6 and 20~cm for a 
synchrotron model of \object{9~Sgr}. The adopted parameters are
$T_{\rm eff}=43\,000$~K, $R_*= 16~{\rm R_{\sun}}$, $d=1.58$~kpc, 
$v_\infty=2\,950~{\rm km\,s^{-1}}$, 
$\dot{M}=2.4\times10^{-6}~{\rm M_{\sun}\,yr^{-1}}$, and 
$L_*=8.1\times 10^5~{\rm L_{\sun}}$. The velocity jump and compression 
ratio were taken from Fig.~\ref{fig:hydro:chi}. The model does not fit 
the observations, because the theoretical synchrotron spectrum between 6 
and 20 cm is essentially flat, contrary to the observed negative spectral 
index. In this sense the same problems as for \object{Cyg OB2 No. 9} appear, 
but now at larger wavelengths.

\section{Conclusions}\label{sect:conclusions}
We have included the radial decline of the shock strength, as predicted by
hydrodynamical models, in our synchrotron-emission model for single O stars, 
and applied it to the widely presumed single O star \object{Cyg~OB2 No.~9}, 
which is a synchrotron emitter. We find that the decline of the shock strength 
no longer allows a wind-embedded shock model to reproduce the observed negative 
spectral index of non-thermal emission. We have investigated a number of 
``escape routes" by which the hydrodynamical predictions might be reconciled 
with the radio observations. We find that a negative spectral index can be
produced by a slower decline of the compression ratio and the velocity jump, 
by the re-acceleration of electrons in many shocks or by adopting a lower 
mass-loss rate. However, none of these escape routes are physically very 
plausible. In particular, re-acceleration by feeding an electron distribution 
through a number of shocks, is in contradiction with modern hydrodynamical 
calculations. This leads us to the conclusion that there is probably no viable 
way to produce the observed spectrum based on hydrodynamical simulations. 

Our model obviously has its limitations. Most important, perhaps, is the
use of one-dimensional (1D) hydrodynamical simulations. Two-dimensional
simulations have been presented for instability-generated structure in the inner
wind (Dessart \& Owocki \cite{DO03}). Overall, these models are characterised 
by a reduced clumpiness of the wind and a larger (radial) velocity dispersion. 
It is not feasible at present to perform 2D-simulations of the wind out to the
distances required for synchrotron-emission models. It cannot be excluded
that such models might show a slower decrease of the shock strength. It is to
be expected though that as radiative driving becomes negligible at distances 
of a few tens of stellar radii (Runacres \& Owocki \cite{RO02}), both the 
compression ratio and the velocity dispersion should eventually decrease 
even in a 2D model. Furthermore, 2D and 3D hydrodynamics allows for many 
additional instabilities (such as Rayleigh-Taylor) to disrupt the structure. 
It is impossible at present to make even an educated guess what a structure 
of such complexity might mean for the efficiency of synchrotron emission.

Based on the current hydrodynamical simulations, which admittedly are
open to question in their details, we suspect that the observed
non-thermal radio emission from O stars cannot be explained by
wind-embedded shocks associated with the instability of the
line-driving mechanism.  The most plausible alternative to emission
by wind-embedded shocks is emission from shocks associated with
colliding stellar winds.  That would imply that all non-thermal O
stars should be members of a binary or multiple system. 

This conclusion is supported by the recent finding (Van Loo \cite{SVLPhD}, 
Blomme et al. \cite{B05}, Rauw et al. \cite{R05}), based on both period analyses 
of radio light curves and spectroscopic evidence, that three presumably single 
stars with non-thermal radio emission are probably binaries. Continued 
systematic radio and spectroscopic monitoring  may lead to the same conclusion 
for the remaining non-thermal radio emitters that are presumed to be single
O stars.

\begin{acknowledgements}

 We thank Sean Dougherty, Julian Pittard and Stan Owocki for carefully 
reading the manuscript and providing valuable comments. 
We thank the referee, R. L. White, for a constructive report that helped 
improve the paper. We also thank 
Achim Feldmeier for useful discussions. SVL gratefully acknowledges a doctoral 
research grant by the Belgian Federal Science Policy Office (Belspo). 
Part of this research was carried out in the framework of the project 
IUAP P5/36 financed by Belspo.
\end{acknowledgements}

\appendix
\section{Effect of clumping on the free-free optical depth}
\label{sect:effect clumping}
The free-free optical depth is usually calculated assuming a smooth wind
(Wright \& Barlow \cite{WB75}) and the effect of density inhomogeneities 
(clumping) is neglected. However, the line-deshadowing instability packs 
the stellar wind material together in dense clumps 
(e.g. Runacres \& Owocki \cite{RO02}). In principle, because the free-free 
optical depth is proportional to the density squared, one expects clumping 
to have an influence on the radio photosphere of the stellar wind.  

This is indeed the case when we take a given amount of stellar wind material
and consider the effect of clumping it. Due to the enhanced opacity of the
clumped wind, a given optical depth (integrated from a location in the wind 
out to the observer) is reached closer to the observer, and as a result the 
radio photosphere of the stellar wind is larger.

However, if we compare a smooth wind and a clumped wind that produce the same 
radio flux, the effect of the enhanced optical depth is cancelled exactly by 
the fact that the clumped wind requires a lower mass loss rate to explain the 
observations. If we assume a uniform, random distribution of
optically thin clumps, then we can use the description of clumping proposed 
by Abbott et al.~(\cite{ABC81}). From their Eq.~(9) we see that the 
optical depth has the following dependence on the clumping factor $f_{\rm cl}$ 
and the mass-loss rate 
\begin{equation}
   \tau_{\nu} \propto \dot{M}^2 f_{\rm cl}.
\end{equation}
(A succinct explanation of the clumping factor is given in Runacres \& Owocki
(\cite{RO02})). The emergent radio flux scales as
\begin{equation}
   F_{\nu} \propto \left(\dot{M}^2 f_{\rm cl}\right)^{2/3}
\end{equation}
(Abbott et al.~\cite{ABC81}, their Eq.~(10)). Therefore, a clumped wind and a 
smooth wind that produce the same radio flux will also have the same optical 
depth. Consequently, a given optical depth is reached at the same radius. This 
means that, in the limit of optically thin clumps, clumping has no effect on 
the values we derive for the radio photosphere of the stellar wind.

When individual clumps become optically thick, the Abbott et al. formalism is no longer valid.
In that case, porosity effects can play a r\^ole. Owocki et al. (\cite{OGS04}) show that for 
optically thick blobs the opacity is effectively reduced. Porosity is characterised by an 
additional parameter (the clump size or equivalent) which complicates the relation between 
the clump properties and the radio flux.

\end{document}